\documentclass[preprint,aps,12pt,amsmath,amssymb,showkeys,nofootinbib,tightenlines,superscriptaddress]{revtex4-1}
\usepackage{amssymb}
\usepackage{amsfonts}
\usepackage{latexsym}
\usepackage{pdfsync}
\usepackage[T1]{fontenc} 
\usepackage{epsf,amsfonts}
\usepackage{epsfig}
\usepackage{amsthm}
\usepackage{amsmath}
\usepackage{graphicx}
\usepackage{color}
\usepackage{slashed}
\usepackage{mathrsfs}
\usepackage{float}
\usepackage{color}
\usepackage{esvect}
\usepackage{mathtools}
\usepackage{bbm}

\newcommand{\VEV}[1]{\left\langle #1\right\rangle}

\newcommand{\MeV}{\;\text{MeV}}

\newcommand{\re}{\mathrm{Re}}
\renewcommand\tabcolsep{6.0pt}

\begin{document}

\title{Octet meson spectra and chiral phase diagram in the improved soft-wall AdS/QCD model}

\author{Zhen Fang}\email{zhenfang@hnu.edu.cn}\affiliation{Department of Applied Physics, School of Physics and Electronics, Hunan University, Changsha 410082, China}
\author{Yue-Liang Wu}\email{ylwu@itp.ac.cn}\affiliation{CAS Key Laboratory of Theoretical Physics, Institute of Theoretical Physics, Chinese Academy of Sciences, Beijing 100190, China}\affiliation{International Centre for Theoretical Physics Asia-Pacific (ICTP-AP), University of Chinese Academy of Sciences, Beijing 100049, China}
\author{Lin Zhang}\email{zhanglin@itp.ac.cn}\affiliation{CAS Key Laboratory of Theoretical Physics, Institute of Theoretical Physics, Chinese Academy of Sciences, Beijing 100190, China}\affiliation{School of Physical Sciences, University of Chinese Academy of Sciences, Beijing 100049, China}

\date{\today}

\begin{abstract}
We give a further study on chiral phase diagram in the improved soft-wall AdS/QCD model with $2+1$ flavors. The equations of motion for the octet pseudoscalar, vector and axial-vector mesons are derived to compute the octet meson spectra and relevant decay constants, by which the model parameters are determined. The chemical potential effects on thermal transition of chiral condensate are investigated, which enables us to obtain the chiral phase diagram in the $\mu-T$ plane. We find that the critical end point linking the crossover transition with the first-order phase transition still exists and locates at $(\mu_B, T_c) \simeq (390 \MeV, 145 \MeV)$, which along with the crossover line are consistent with lattice result and experimental analysis from relativistic heavy ion collisions.
\end{abstract}

\keywords{octet meson spectra, chiral transition, chiral phase diagram, AdS/QCD}

\maketitle
\section{Introduction}\label{introduce1}

As an important research area of quantum chromodynamics (QCD), there are still many puzzles on the phase transition of strongly interacting matters. Lattice simulations show evidence that the QCD matters exhibit a crossover transition with the increase of temperature $T$ at zero baryon chemical potential \cite{Aoki:2006we,Bazavov:2011nk,Bhattacharya:2014ara}. It is generally believed that there is a critical end point (CEP) at some finite chemical potential in QCD phase diagram, and the crossover transition converts into first-order phase transition as the baryon chemical potential $\mu_B$ increases beyond the CEP \cite{Fodor:2001pe,Fodor:2004nz,Gavai:2004sd}. The physics of the CEP and the search of it are crucial for our understanding of QCD phase diagram, which have attracted extentive studies for decades, both theoretically and experimentally \cite{Stephanov:2007fk,Luo:2017faz}. However, no final conclusion has been reached on these matters. Lattice QCD as the first-principle calculation is hindered by the sign problem at finite chemical potential, though some tentative methods have been proposed to address this issue \cite{Fodor:2001au,Allton:2002zi,deForcrand:2002hgr,Ding:2017giu}. The QCD phase diagram and the relevant properties of CEP have also been studied in various effective models, see Ref. \cite{Stephanov:2007fk} for a review.

In this work, we continue the holographic QCD program aimed to study the low-energy physics of strong interaction in terms of the anti-de Sitter/conformal field theory (AdS/CFT) correspondence \cite{Maldacena:1997re,Gubser:1998bc,Witten:1998qj}. There have been large amounts of holographic studies on the low-energy hadron physics over the past decades, including hadron spectrum, QCD thermodynamics and many other properties \cite{DaRold:2005mxj,Erlich:2005qh,Karch:2006pv,deTeramond:2005su,Brodsky:2014yha,Babington:2003vm,Kruczenski:2003uq,Sakai:2004cn,Sakai:2005yt,Csaki:2006ji,Cherman:2008eh,Gherghetta:2009ac,Kelley:2010mu,Sui:2009xe,Sui:2010ay,Cui:2013xva,Cui:2014oba, Li:2012ay,Li:2013oda,Policastro:2001yc,Cai:2009zv,Cai:2008ph,Sin:2004yx,Shuryak:2005ia,Nastase:2005rp, Nakamura:2006ih,Sin:2006pv,Janik:2005zt, Herzog:2006gh,Gubser:2008yx,Gubser:2008ny,Gubser:2008sz,Noronha:2009ud,DeWolfe:2010he,Finazzo:2013efa,Finazzo:2014zga,Yaresko:2013tia,Li:2011hp,Li:2014hja,Li:2014dsa,Fang:2016uer,Fang:2016dqm,Chelabi:2015cwn,Chelabi:2015gpc,Fang:2015ytf,Evans:2016jzo,Mamo:2016xco,Fang:2016cnt,Dudal:2016joz,Dudal:2018rki,Ballon-Bayona:2017dvv,Critelli:2017oub,Critelli:2017euk,Critelli:2018osu,Rougemont:2018ivt,ChenXun:2019zjc}. Although AdS/CFT has been taken as a powerful tool in the description of low-energy QCD, a caveat must be given here on the validity of this method in view of the assumption of large N limit and the supergravity approximation that may be invalid in the ultraviolet (UV) region of QCD with asymptotic freedom. The string loop correction is indeed required to provide an adequate account of QCD thermodynamics in the holographic framework. Nevertheless, in the current stage, we still need a down-to-earth treatment in order to understand the holographic picture of low-energy QCD. 

This work focuses on chiral transition in AdS/QCD with $2+1$ flavors, especially on the CEP-related properties of chiral phase diagram. We remark that a more realistic description for QCD phase diagram demands a detailed study on the equation of state, which is intimately related to the dual gravity background \cite{Gubser:2008yx,Gubser:2008ny,Gubser:2008sz,Noronha:2009ud,DeWolfe:2010he,Finazzo:2013efa,Finazzo:2014zga,Yaresko:2013tia,Li:2011hp,Li:2014hja}. Following the previous studies \cite{Colangelo:2011sr,Jarvinen:2011qe,Alho:2012mh,Alho:2013hsa,Jarvinen:2015ofa,Li:2016smq,Bartz:2016ufc,Bartz:2017jku,Li:2017ple,Chen:2018msc}, we just adopt a fixed black hole background to study the thermodynamical behaviors of chiral transition that comes from the flavor sector of AdS/QCD. We expect that the sensible chiral transition behaviors obtained from our model can still be generated somehow from AdS/QCD with dynamical background.

In Ref. \cite{Fang:2016nfj}, we proposed an improved soft-wall AdS/QCD model with linear confinement and spontaneous chiral symmetry breaking in the two-flavor case, where the light meson spectra and the properties of chiral transition have been studied in detail. We then generalized this two-flavor AdS/QCD model to the $2+1$ flavor case, where a 't Hooft determinant term of the bulk scalar field turns out to be crucial for the description of quark-mass phase diagram \cite{Fang:2018vkp}. The chemical potential effects on chiral transition have also been studied in this model, where the chiral phase diagram in the $\mu -T$ plane can be obtained with a CEP linking the crossover transition with the first-order phase transition \cite{Fang:2018axm}. However, the model parameters in the aforementioned works are all tuned artificially to generate correct chiral transition behaviors, without any constraint from low-energy hadron properties. In this work, we try to compute the octet meson spectra and the related decay constants in the improved soft-wall AdS/QCD model with $2+1$ flavors, by which the model parameters can be constrained and the relevant properties of chiral transition at finite baryon chemical potential can be investigated.

The paper is organized as follows. In Sec. \ref{model}, we outline the improved soft-wall AdS/QCD model with $2+1$ flavors. In Sec. \ref{mass-octet}, we first derive the equation of motion (EOM) of octet pseudoscalar, vector and axial-vector mesons, and then fit the model parameters by the octet meson spectra and the related decay constants. In Sec. \ref{chiraltran}, we investigate the chiral transition behaviors at finite $\mu_B$, from which the chiral phase diagram in the $\mu-T$ plane can be obtained. In Sec. \ref{conclution}, we come to a brief summary of our work and conclude with some remarks.

\section{The improved soft-wall AdS/QCD model with $2+1$ flavors}\label{model}

The improved soft-wall AdS/QCD model is constructed on the background of AdS$_5$ spacetime with the metric ansatz
\begin{equation}\label{metric}
ds^2=e^{2A(z)}\left(\eta_{\mu\nu}dx^{\mu}dx^{\nu}-dz^2\right) ,
\end{equation}
where $\eta_{\mu\nu}=(+1,-1,-1,-1)$ and $A(z)=-\mathrm{log}\frac{z}{L}$ (the AdS radius is set to be $L=1$ for simplicity).  

The bulk action of the improved soft-wall AdS/QCD model with $2+1$ flavors can be written as
\begin{equation}\label{2+1-act}
S =\int d^{5}x\,\sqrt{g}\,e^{-\Phi(z)}\left[\mathrm{Tr}\{|DX|^{2}-m_5^2(z)|X|^{2}
-\lambda |X|^{4}-\frac{1}{4g_{5}^2}(F_{L}^2+F_{R}^2)\} -\gamma\,\re\{\det X\}\right] ,
\end{equation}
where the covariant derivative of bulk scalar field has the form $D^MX=\partial^MX-i A_L^MX+i X A_R^M$, and the field strength $F_{L,R}^{MN}=\partial^MA_{L,R}^N-\partial^NA_{L,R}^M-i[A_{L,R}^M,A_{L,R}^N]$ with the chiral gauge field $A_{L,R}^M=A_{L,R}^{a,M}T^a =\frac{1}{2}A_{L,R}^{a,M}\lambda^a$, where $\mathrm{Tr}(T^aT^b)=\frac{1}{2}\delta^{ab}$ and $\lambda^a$ denote Gell-Mann matrices. The gauge coupling is $g_5^2=12\pi^2/N_c$ with $N_c$ being the color number \cite{Erlich:2005qh}. The running mass of bulk scalar field takes the form $m_5^2(z)=-3-\mu_c^2\,z^2$ with the constant term $-3$ determined by the mass-dimension relation in AdS/CFT \cite{Erlich:2005qh} and the infrared (IR) asymptotics of $m_5^2(z)$ empirically related to the low-energy hadron properties \cite{Fang:2016nfj}. To give a better fitting for the $\rho$ meson spectrum, we will use a modified dilaton field in this work,
\begin{equation}\label{Phi}
\Phi(z) =\mu_{g}^2\,z^2\left(1-e^{-\frac{1}{4}\mu_{g}^2z^2}\right),
\end{equation}
which has the IR limit $\Phi(z\to\infty)\sim z^2$ to reproduce the Regge spectra of highly excited mesons. Here we remark that the ultraviolet (UV) asymptotics of $\Phi(z)$ has little effects on the properties of chiral phase diagram and other meson spectra in our framework. The 't Hooft determinate term $\re\{\det X\}$ has been introduced into the bulk action for the correct realization of chiral transition in the $2+1$ flavor case \cite{Chelabi:2015gpc}.

The vacuum expectation value (VEV) of bulk scalar field takes the form
\begin{equation}\label{VEVs}
\langle X \rangle=\frac{1}{\sqrt{2}}
\begin{pmatrix}
\chi_u(z) & 0 & 0 \\
0 & \chi_d(z) & 0 \\
0 & 0 & \chi_s(z)
\end{pmatrix}
\end{equation}
with $\chi_u=\chi_d$ for the $2+1$ flavor case. The action of the scalar VEV $\VEV{X}$ can be obtained from the bulk action (\ref{2+1-act}) as
\begin{equation}\label{2+1-vev-act}
S_{\chi} =\int d^{5}x\,\sqrt{g}\,e^{-\Phi(z)}\left[\mathrm{Tr}\{\partial^{z}\VEV{X}\partial_{z}\VEV{X} -m_5^2(z)\VEV{X}^{2}
-\lambda\VEV{X}^{4}\} -\gamma\,\det\VEV{X}\right],
\end{equation}
from which the EOMs of $\chi_{u}$ and $\chi_{s}$ can be derived as
\begin{align}
\chi_{u}'' +\left(3A'-\Phi'\right)\chi'_{u} -e^{2A}\left(m_5^2\chi_u +\lambda\chi_u^3 +\frac{\gamma}{2\sqrt{2}}\chi_u\chi_s \right)  &=0,   \label{vevX-eom1}  \\
\chi_{s}'' +\left(3A'-\Phi'\right)\chi'_{s} -e^{2A}\left(m_5^2\chi_s +\lambda\chi_s^3 +\frac{\gamma}{2\sqrt{2}}\chi_{u}^{2}\right)  &=0,   \label{vevX-eom2}
\end{align}
which are coupled with each other as a result of the 't Hooft determinate term of the bulk scalar field.

The UV asymptotic forms of the scalar VEV $\chi_{u,s}$ near the boundary can be obtained from Eqs. (\ref{vevX-eom1}) and (\ref{vevX-eom2}) as
\begin{align}
\chi_u(z \sim 0) =\frac{1}{\sqrt{2}}&\left[ m_u\,\zeta\,z-\frac{1}{4}\,m_u\, m_s\,\gamma \, \zeta^2\,z^2+\frac{\sigma_u}{\zeta}z^3 +\frac{1}{16} m_{u}\zeta\left(-\frac{1}{2}\, m_{s}^{2}\, \gamma^{2}\,\zeta^{2} \right.\right.  \nonumber \\
&\quad\left.\left.-\frac{1}{2}\,m_{u}^{2}\,\gamma^{2}\,\zeta^{2}+4\,m_{u}^{2}\,\zeta^{2}\,\lambda-8\,\mu_{c}^{2} \right)\,z^{3}\, \log{z}+\cdots\right] ,      \label{asy-chiu1}  \\
\chi_s(z \sim 0) =\frac{1}{\sqrt{2}}&\left[m_s\,\zeta\,z-\frac{1}{4}\,m_u^2\,\gamma \, \zeta^2\,z^2+\frac{\sigma_s}{\zeta}z^3 +\frac{1}{16} m_{s}\zeta\left(-\,m_{u}^{2}\, \gamma^{2}\,\zeta^{2}\right.\right.  \nonumber \\
&\quad \left.+4\,m_{s}^{2}\,\zeta^{2}\,\lambda-8\,\mu_{c}^{2}\right)
\,z^{3}\, \log{z} +\cdots\bigg] ,      \label{asy-chis1} 
\end{align}
where $m_{u,s}$ denote current quark masses and $\sigma_{u,s}$ denote chiral condensates, and the normalization constant $\zeta$ is fixed as $\zeta=\frac{\sqrt{N_c}}{2\pi}$ \cite{Cherman:2008eh}. The coefficient $\frac{1}{\sqrt{2}}$ is necessary to attain the Gell-Mann–Oakes–Renner (GOR) relation $m_{\pi}^2f_\pi^2 =2m_u\sigma_u$ for the two-flavor case \cite{Erlich:2005qh}. The IR asymptotics of scalar VEV should take the linear form $\chi_{u,s}(z\to\infty) \sim z$ to generate the mass split of chiral partners \cite{Fang:2016nfj}. With the above UV asymptotic forms and the IR boundary conditions, the scalar VEV $\chi_{u,s}$ can be solved numerically from Eqs. (\ref{vevX-eom1}) and (\ref{vevX-eom2}), and the chiral condensates $\sigma_{u,s}$ can be extracted.

\section{Octet meson spectra and decay constants}\label{mass-octet}

\subsection{Input parameters}

Now we consider the octet meson spectra and the related decay constants, which will be used to constrain the model parameters $\mu_{g}$, $\gamma$, $\lambda$ and $\mu_{c}$. The parameter $\mu_{g}$ can be fixed by the $\rho$ meson spectrum, while the parameters $\gamma$ and $\lambda$ can be determined by $\pi$ meson spectrum and pion decay constant. The last parameter $\mu_c$ is strongly correlated with the chemical potential value of CEP, and it also affects the global fitting of octet meson spectra. The physical quark masses $m_{u,s}$ are slightly tuned within the error range of experimental data in order to attain the best fitting for the ground-state masses of octet pseudoscalar mesons. 

We first derive the EOMs of octet pseudoscalar, vector and axial-vector mesons from the linearized action of relevant bulk fields, and then we compute the mass spectra of these octet mesons and related decay constants. The fixed parameter values are listed in Table \ref{parameter-fit}.
\begin{table}
\begin{center}
\begin{tabular}{cccccc}
\hline\hline
$m_u$(MeV) &  $m_s$(MeV) & $\mu_{g}$(MeV) & $\mu_c$(MeV)  & $\lambda$ & $\gamma$ \\
\hline
 3.24    &    98    &    480     &     877.8     &    130   &   -69.53  \\
\hline\hline
\end{tabular}
\caption{The input values of model parameters in the numerical calculation. The quark masses $m_{u,s}$ are taken from Ref. \cite{Tanabashi:2018oca}.}
\label{parameter-fit}
\end{center}
\end{table}

\subsection{Octet pseudoscalar mesons}

Following the usual procedure, the bulk scalar field $X$ can be decomposed into
\begin{equation}\label{X-decomp}
X=\xi(X_{0}+S^{a}T^a +S^0T^0)\xi, \quad  \xi=\mathrm{exp}(iT^a\pi^a),
\end{equation}
where $\pi^a$ are pseudoscalar fields and $S^a$ ($S^0$) are $SU(3)$ octet (singlet) scalar fields. We will neglect the scalar part in our work. In the axial gauge $A_z=0$, the axial gauge fields can be written as $A_{\mu}^a=A_{\mu \bot}^a+\partial_{\mu}\phi^a$ to eliminate the cross terms of the pseudoscalar and axial-vector fields. With the Kaluza-Klein (KK) decomposition $\pi^a(x,z)=\sum_n\varphi_n(x)\pi^a_n(z)$, the EOMs of octet pseudoscalar mesons can be derived as
\begin{align}
\partial_z\left(e^{A-\Phi}\partial_z\phi^a_n\right) +2g_{5}^2e^{3A-\Phi}(M^2_A)_{ab} \left(\pi^b_n -\phi^b_n\right)  &=0 ,     \label{3f-PS-eom2-2}  \\
m_n^2\partial_z\phi^a_n -2g_{5}^2 e^{2A}(M^2_A)_{ab}\partial_z\pi^b_n  &=0     \label{3f-PS-eom4-2}
\end{align}
with
\begin{align}\label{M-A2}
M^2_A =\begin{pmatrix} \chi_u^2 \mathbf{1}_{3\times3}  & 0 & 0 \\ 0 & \frac{1}{4}(\chi_u+\chi_s)^2\mathbf{1}_{4\times4} & 0 \\ 0 & 0 & \frac{1}{3}(\chi_u^2+2\chi_s^2) \end{pmatrix},
\end{align}
which can be solved numerically with the boundary condition $\pi_n^a(z\to 0) =\phi_n^a(z\to 0) =\partial_z\phi_n^a(z\to\infty) =0$. 

The computed mass spectra of octet pseudoscalar mesons are presented in Table \ref{pi-spectrum1}, where the experimental data are also shown for comparison. The data choosing is based on the suggested quark model assignments for the observed light mesons \cite{Tanabashi:2018oca}.
\begin{table}
\renewcommand\tabcolsep{8.0pt}
\begin{center}
 \begin{tabular}{ccccccc}
  \hline\hline
 $n$ & $\pi$ exp. (MeV) & Model  & $K$  exp. (MeV)  & Model & $\eta$  exp. (MeV)  &  Model  \\
   \hline\hline
  0 & $139.57$ & $139.57$ & $493.677\pm0.016$ & $492.85$ & $547.862\pm0.017$ & $585.19$ \\
   \hline
  1 & $1300\pm100$ & $1447$ & $1460$ & $1472$ & $1476\pm4$ & $1485$ \\
   \hline
  2 & $1812\pm12$ & $1817$ & $1874\pm43$ & $1836$ & $1751\pm15$ & $1846$ \\
   \hline\hline
     \end{tabular}
\caption{The model results of the mass spectra of octet pseudoscalar mesons, which are compared with experimental data taken from the suggested quark-model assignment for the observed light mesons \cite{Tanabashi:2018oca}.}
\label{pi-spectrum1}
\end{center}
\end{table}

\subsection{Octet vector and axial-vector mesons}

The chiral gauge fields are recombined into the vector field $V^M=\frac{1}{2}(A_L^M+A_R^M)$ and the axial-vector field $A^M=\frac{1}{2}(A_L^M-A_R^M)$. In the axial gauge $V_z= A_z=0$ and with the KK decomposition, the EOMs of octet vector and axial-vector mesons can be derived from the bulk action (\ref{2+1-act}) as
\begin{align}
\partial_z\left(e^{A-\Phi}\partial_zV_n^a\right) -2g_{5}^2e^{3A-\Phi}(M^2_V)_{ab}V_n^b +m_{V^a_n}^2e^{A-\Phi}V_n^a &=0,   \label{V-eom1} \\
\partial_z\left(e^{A-\Phi}\partial_zA_n^a\right) -2g_{5}^2e^{3A-\Phi}(M^2_A)_{ab}A_n^b +m_{A^a_n}^2e^{A-\Phi}A_n^a &=0,   \label{A-eom1}
\end{align}
where the matrix $M^2_A$ is given in (\ref{M-A2}) and $M^2_V$ has the form
\begin{align}\label{M-V1}
M^2_V =\begin{pmatrix} \mathbf{0}_{3\times3} & 0 & 0 \\ 0 & \frac{1}{4}(\chi_u-\chi_s)^2\mathbf{1}_{4\times4} & 0 \\ 0 & 0 & 0 \end{pmatrix}.
\end{align}
Note that the octet axial-vector mesons are incorporated in the transverse part of axial gauge fields $A_{\mu \bot}^a$.

In terms of the redefinitions $V^a_n=e^{\omega/2} v^a_n$ and $A^a_n=e^{\omega/2} a^a_n$ with $\omega=\Phi -A$, the Eqs. (\ref{V-eom1}) and (\ref{A-eom1}) can be transformed into the Schr$\ddot{o}$dinger form
\begin{align}
\partial_z^2v_n^a+\left(\frac{1}{2}\omega'' -\frac{1}{4}\omega'^2\right)v^a_n -2g_{5}^2e^{2A}(M^2_V)_{ab}v_n^b +m_{V^a_n}^2 v^a_n &=0,   \label{V-eom2} \\
\partial_z^2a_n^a+\left(\frac{1}{2}\omega'' -\frac{1}{4}\omega'^2\right)a^a_n -2g_{5}^2e^{2A}(M^2_A)_{ab}a_n^b +m_{A^a_n}^2 a^a_n &=0.   \label{A-eom2}
\end{align}
The mass spectra of octet vector and axial-vector mesons can be obtained by solving the eigenvalue problem of Eqs. (\ref{V-eom2}) and (\ref{A-eom2}) with the following boundary condition
\begin{align}
& v_n^a(z\to 0)=0, \quad  \partial_zv_n^a(z\to \infty) =0;   \label{V-bound} \\
& a_n^a(z\to 0)=0, \quad  \partial_za_n^a(z\to \infty) =0.   \label{A-bound}
\end{align}

The numerical results are presented in Table \ref{rho-spectrum1} and \ref{a1-spectrum1}, where only the data of ground-state mesons in the axial-vector octet have been shown in consideration of experimental uncertainty \cite{Tanabashi:2018oca}. We find that the mass spectra of $\phi$ and $f_1$ computed from the model have large discrepancies with experimental values. This is due to the fact that the EOMs of isovector and isosinglet states in the vector octet are of the same form as a result of the specific $M_V^2$, while the flavor symmetry breaking terms built in $M_A^2$ differ so little from each other that cannot distinguish the masses of different axial-vector mesons. One might introduce high-order terms into the bulk action to improve the model results of vector and axial-vector meson spectra without affecting the pseudoscalar part, e.g., $D_{[M}X D_{N]}X^{\dagger}F_L^{MN}$, $D_{[M}X^{\dagger}D_{N]}XF_R^{MN}$, $XX^{\dagger}F_L^{MN}F_{LMN}$, $X^{\dagger}XF_R^{MN}F_{RMN}$, $XF_R^{MN}X^{\dagger}F_{LMN}$. However, in this way, the predictive power would be decreased with more free parameters introduced. Thus we will not consider it in this work. One can refer to Ref. \cite{Sui:2010ay} for the effects of high-order terms on the octet meson spectra. 
\begin{table*}
\renewcommand\tabcolsep{8.0pt}
\begin{center}
 \begin{tabular}{cccccccc}
  \hline\hline
 $n$ & $\rho$  exp. (MeV)  &  Model & $K^{*}$  exp. (MeV)  & Model & $\phi$  exp. (MeV)  & Model \\
   \hline\hline
  0 & $775.26\pm0.25$ & $775.1$ & $891.76\pm0.25$ & $775.2$ & $1019.461\pm0.016$  & $775.1$ \\
   \hline
  1 & $1465\pm25$ & $1335$ & $1421\pm9$ & $1336$ & $1680\pm20$  & $1335$ \\
   \hline
  2 & $1720\pm20$ & $1714$ & $1718\pm18$ & $1714$ & $2188\pm10$  & $1714$ \\
   \hline\hline
     \end{tabular}
\caption{The model results of the mass spectra of octet vector mesons, which are compared with experimental data taken from the suggested quark-model assignment for the observed light mesons \cite{Tanabashi:2018oca}.}
\label{rho-spectrum1}
\end{center}
\end{table*}
\begin{table*}
\renewcommand\tabcolsep{8.0pt}
\begin{center}
 \begin{tabular}{cccccccc}
  \hline\hline
 $n$  &  $a_1$ exp.(MeV)  &  Model  &  $K_1$ exp.(MeV)  &  Model  &  $f_1$ exp.(MeV)  &  Model  \\
   \hline\hline
  0 & $1230\pm40$ & $1115$ & $1272\pm7$ & $1121$ & $1426.4\pm0.9$ & $1122.8$  \\
   \hline
  1 & $\cdots$ & $1525$ & $\cdots$ & $1528$ & $\cdots$  & $1530$ \\
   \hline
  2 & $\cdots$ & $1854$ & $\cdots$ & $1856$ & $\cdots$  & $1857$ \\
   \hline\hline
\end{tabular}
\caption{The model results of the mass spectra of octet axial-vector mesons, which are compared with experimental data taken from the suggested quark-model assignment for the observed light mesons \cite{Tanabashi:2018oca}.}
\label{a1-spectrum1}
\end{center}
\end{table*}

\subsection{Decay constants of the pseudoscalar and (axial-)vector mesons}

According to AdS/CFT \cite{Erlich:2005qh}, the decay constants of octet pseudoscalar mesons can be extracted from the two-point correlation functions of axial-vector currents,
\begin{align}\label{pi-decay}
f_{\pi^a}^2  &=-\frac{1}{g_5^2}e^{A-\Phi}\partial_z A^a(0,z)|_{z\to0},
\end{align}
where $A^a(0,z)$ is the solution of Eq. (\ref{A-eom1}) with $m_{A^a_n}=0$ and the boundary condition $A^a(0,0)=1$ and $\partial_z A^a(0,\infty)=0$. It should be noted that we have taken the limit $m_{\pi^a}\to 0$ in the derivation of the decay constants $f_{\pi^a}$, which is only a suitable approximation for $\pi$ meson \cite{Erlich:2005qh}.
The decay constants of $\rho$ and $a_1$ mesons are given by
\begin{align}\label{rho-decay}
F_{\rho}^2 &=\frac{1}{g_5^2}\left(e^{A-\Phi}\partial_z V_{\rho}(z)|_{z\to0}\right)^2,\\
F_{a_1}^2  &=\frac{1}{g_5^2}\left(e^{A-\Phi}\partial_z A_{a_1}(z)|_{z\to0}\right)^2,
\end{align}
where $V_{\rho}(z)$ and $A_{a_1}(z)$ are the ground-state wave functions of $\rho$ and $a_1$ normalized by $\int dz\,e^{A-\Phi}V_{\rho}^2 =\int dz\,e^{A-\Phi}A_{a_1}^2 =1$.
We show the model results of the decay constants $f_{\pi}$, $f_{K}$, $F_{\rho}$ and $F_{a_1}$ along with the experimental values in Table \ref{decay-const}, where $f_{\pi}$ is taken as an input to determine the values of $\gamma$ and $\lambda$.
\begin{table}
\begin{center}
\begin{tabular}{ccccc}
\hline\hline
     &  $f_{\pi}$(MeV) &  $f_{K}$(MeV) & $F_{\rho}^{1/2}$(MeV)  & $F_{a_1}^{1/2}$(MeV) \\
\hline
 Exp. &  $92.4$  &  $110$  &  $346.2\pm1.4$  &  $433\pm13$  \\
 \hline
Model  &   $92.4$   &   100     &     307.1      &    350  \\
\hline\hline
\end{tabular}
\caption{The model results and the experimental values of related decay constants \cite{Erlich:2005qh,Tanabashi:2018oca}.}
\label{decay-const}
\end{center}
\end{table}

\section{Chiral transition and phase diagram}\label{chiraltran}

With the model parameters fixed by the octet meson spectra and the relevant decay constants, we now study the chemical potential effects on chiral transition in the improved soft-wall AdS/QCD model following Ref. \cite{Fang:2018axm}. To introduce temperature $T$ and baryon chemical potential $\mu_B$, we adopt the AdS/Reissner-Nordstrom (AdS/RN) black hole as the bulk background in terms of the metric ansatz
\begin{equation}\label{AdSRN1}
ds^2=e^{2A(z)}\left(f(z)dt^2-dx^{i\,2}-\frac{dz^2}{f(z)}\right)
\end{equation}
with
\begin{align}\label{AdSRN2}
f(z) &=1-(1+Q^2)\left(\frac{z}{z_h}\right)^4 +Q^2\left(\frac{z}{z_h}\right)^6,
\end{align}
where the charge $Q=\mu_q z_h$ with $z_h$ being the event horizon of black hole and $\mu_q$ being the quark chemical potential that is related to the baryon chemical potential by $\mu_B=3\mu_q$ \cite{Colangelo:2011sr}. The Hawking temperature is given by the formula
\begin{align}\label{T1}
T =\frac{1}{4\pi}\left|\frac{df}{dz}\right|_{z_{h}} =\frac{1}{\pi z_h}\left(1-\frac{Q^2}{2}\right)
\end{align}
with $0<Q<\sqrt{2}$.

In terms of the metric ansatz (\ref{AdSRN1}) of the AdS/RN black hole, the EOMs of the scalar VEV $\chi_{u,s}$ can be derived as
\begin{align}
\chi_{u}'' +\left(\frac{f'}{f}+3A'-\Phi'\right)\chi'_{u} -\frac{e^{2A}}{f}\left(m_5^2\chi_u +\lambda\chi_u^3 +\frac{\gamma}{2\sqrt{2}}\chi_u\chi_s \right)  &=0,   \label{vevX-eomT1}  \\
\chi_{s}'' +\left(\frac{f'}{f}+3A'-\Phi'\right)\chi'_{s} -\frac{e^{2A}}{f}\left(m_5^2\chi_s +\lambda\chi_s^3 +\frac{\gamma}{2\sqrt{2}}\chi_{u}^{2}\right)  &=0.   \label{vevX-eomT2}
\end{align}
By the prescription of AdS/CFT \cite{Erlich:2005qh}, the UV asymptotic forms of $\chi_{u,s}$ can be obtained from the above EOMs with the chiral condensates $\sigma_{u,s}$ now depending on $\mu_B$ and $T$. The differences from those in Eqs. (\ref{asy-chiu1}) and (\ref{asy-chis1}) are only incorporated in the high-order terms of $z$. At finite $\mu_B$ and $T$, the Eqs. (\ref{vevX-eomT1}) and (\ref{vevX-eomT2}) admit natural boundary conditions on the horizon $z=z_h$ imposed by the regular properties of $\chi_{u,s}$,
\begin{align}
\left.f'\chi'_{u} -e^{2A}\left(m_5^2\chi_u +\lambda\chi_u^3 +\frac{\gamma}{2\sqrt{2}}\chi_u\chi_s \right)\right|_{z=z_{h}}  &=0,   \label{BC1}  \\
\left.f'\chi'_{s} -e^{2A}\left(m_5^2\chi_s +\lambda\chi_s^3 +\frac{\gamma}{2\sqrt{2}}\chi_{u}^{2}\right)\right|_{z=z_{h}}  &=0.   \label{BC2}
\end{align}

\begin{figure}
\begin{center}
\includegraphics[width=68mm,clip=true,keepaspectratio=true]{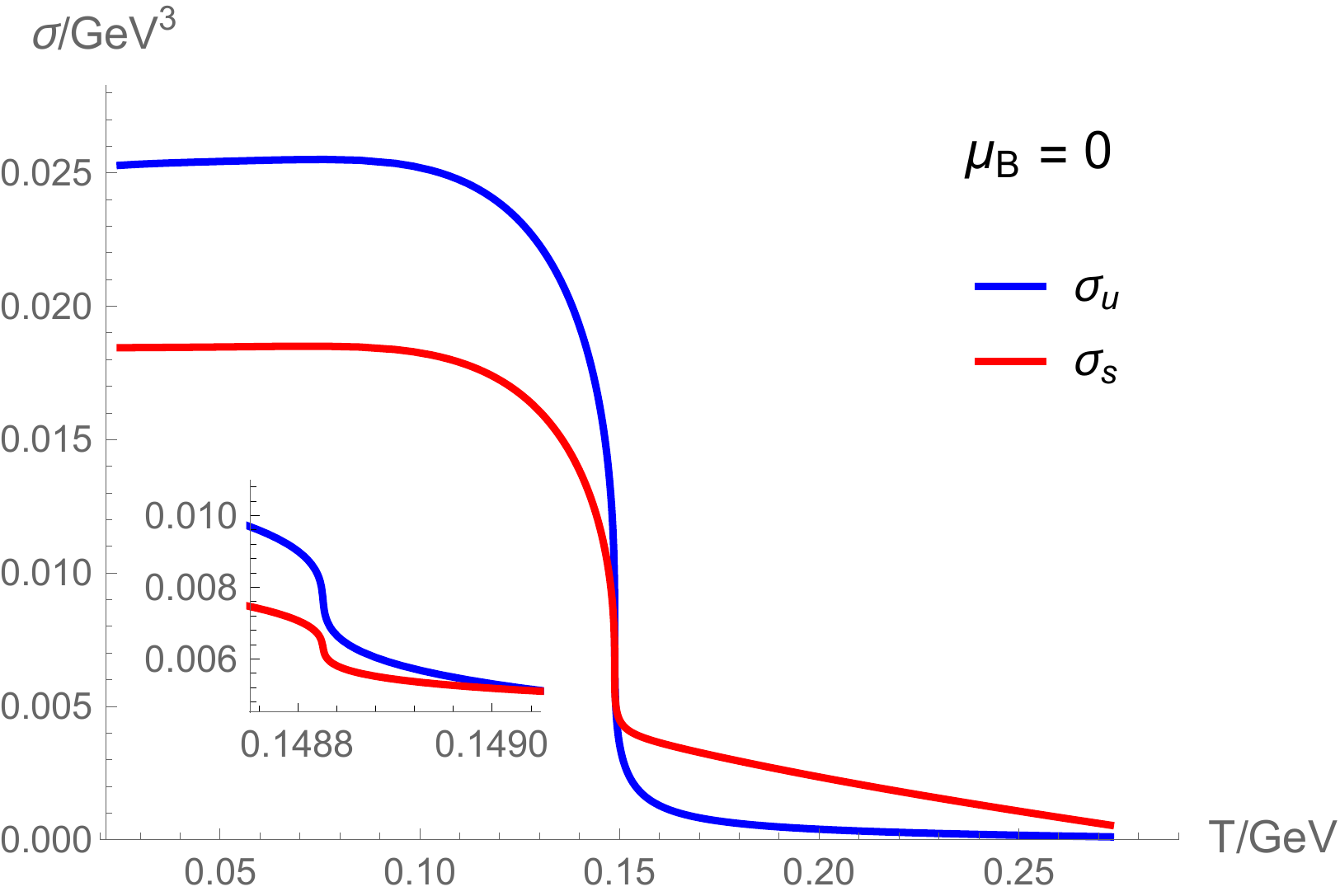}
\hspace*{0.6cm}
\includegraphics[width=68mm,clip=true,keepaspectratio=true]{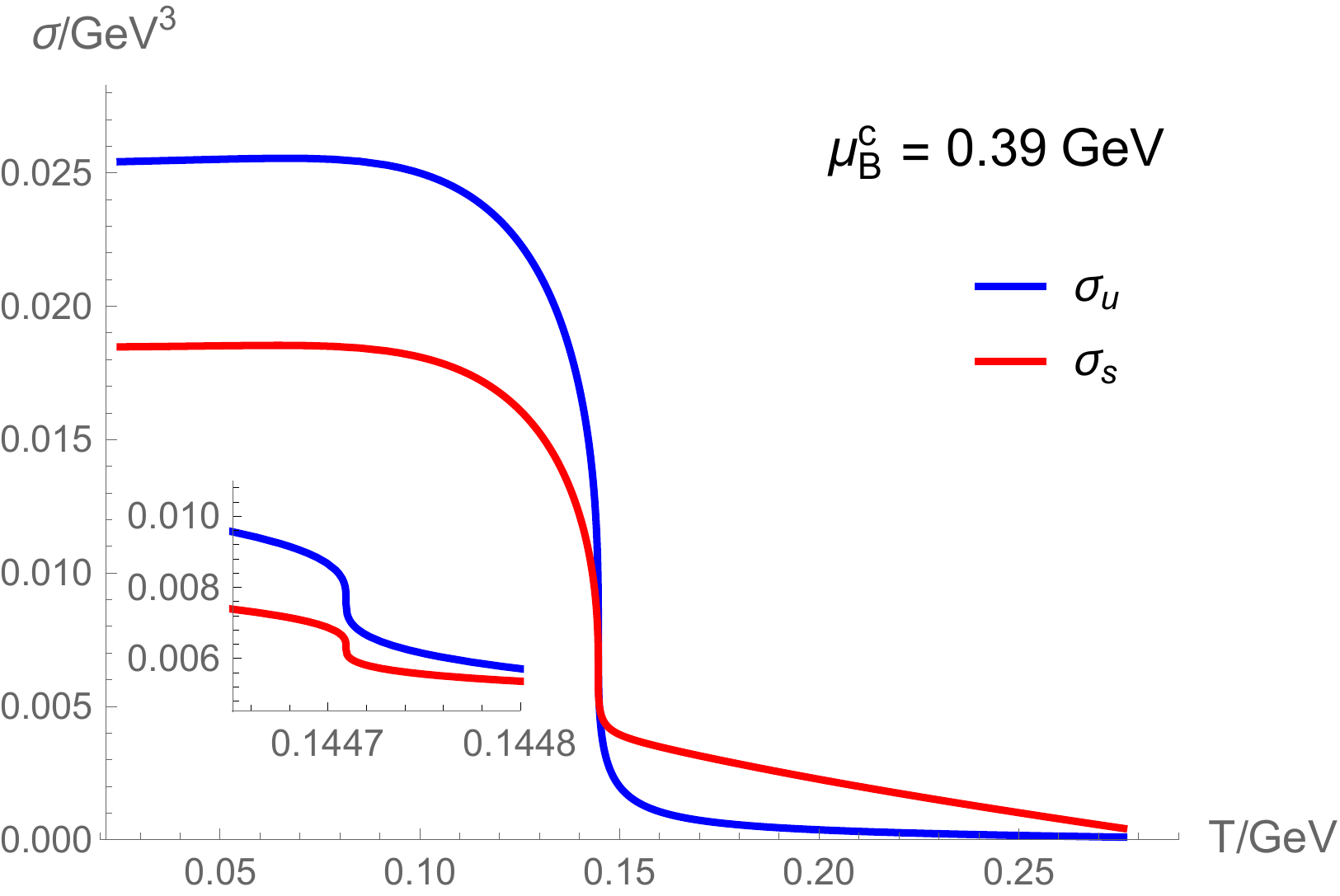}
\vspace{0.35cm} \\ 
\includegraphics[width=68mm,clip=true,keepaspectratio=true]{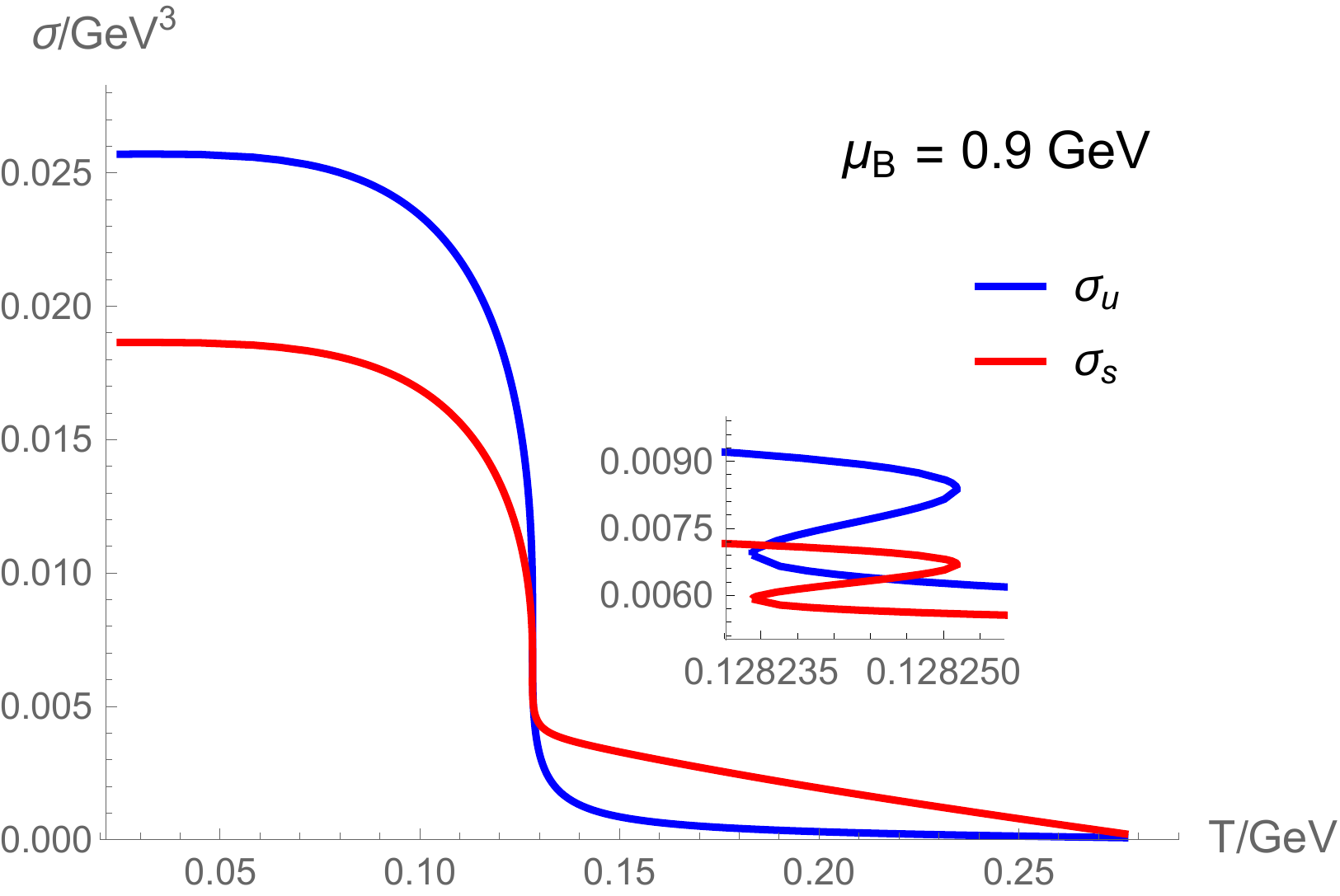}
\hspace*{0.6cm}
\includegraphics[width=68mm,clip=true,keepaspectratio=true]{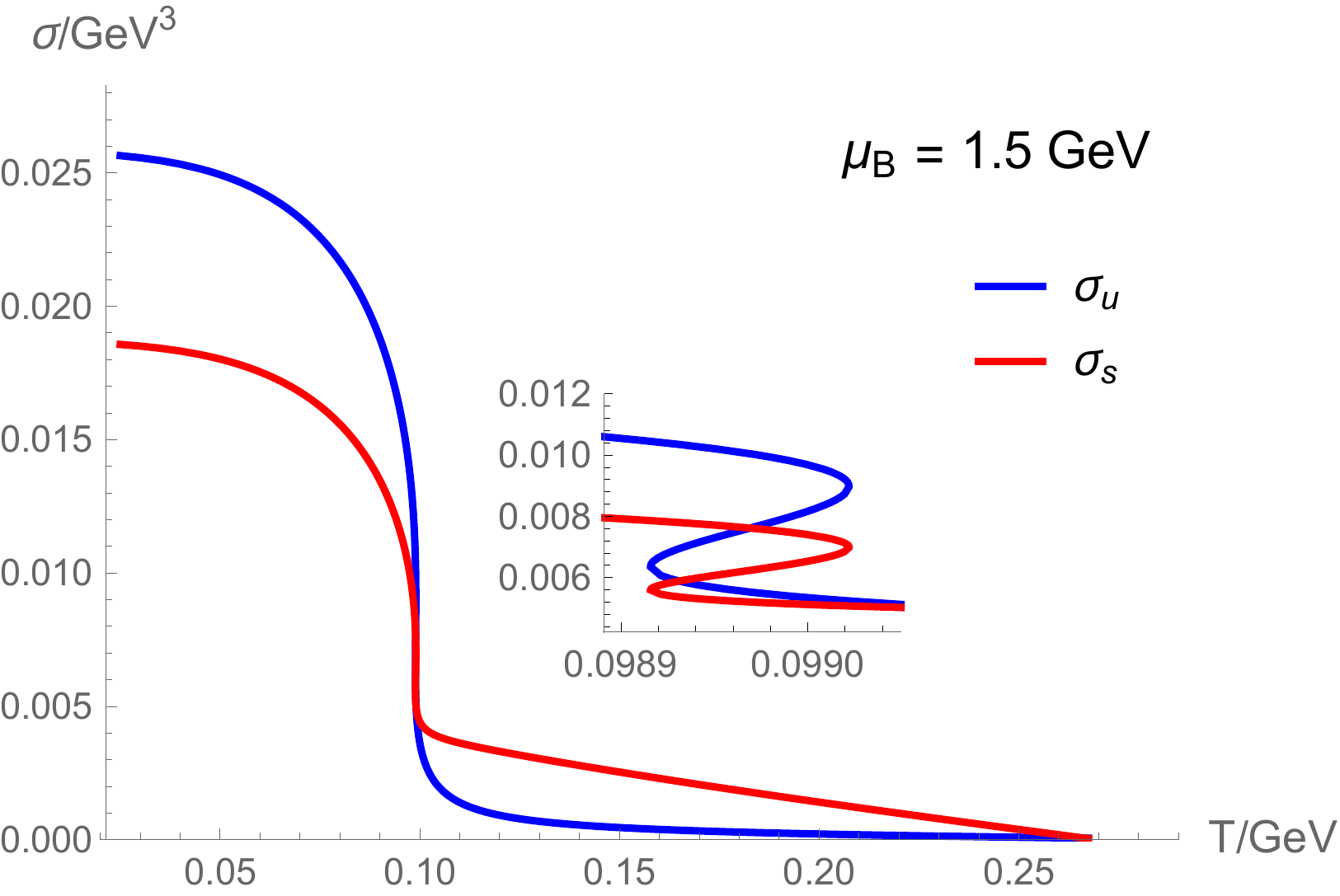}
\vskip -1cm \hskip 0.7 cm
\end{center}
\caption{The chiral transition behaviors of the condensates $\sigma_{u}$ and $\sigma_{s}$ with the temperature $T$ at $\mu_B=0, 0.39, 0.9, 1.5\,\mathrm{GeV}$.}
\label{sigma-T-mu}
\end{figure}

As in Ref. \cite{Fang:2018vkp}, we can solve Eqs. (\ref{vevX-eomT1}) and (\ref{vevX-eomT2}) numerically with the given boundary conditions to obtain the profiles of scalar VEV $\chi_{u,s}$ and thus the chiral condensates $\sigma_{u,s}$ as functions of $\mu_B$ and $T$. The thermal transitions of $\sigma_{u,s}$ with temperature $T$ at four different chemical potentials are shown in Fig. \ref{sigma-T-mu}, where the distinction between the behaviors of $\sigma_{u}(T)$ and that of $\sigma_{s}(T)$ is due to the mass difference of $m_u$ and $m_s$. We can see that the chiral transition is a crossover at small $\mu_B$, and eventually turns into a first-order phase transition with the increase of $\mu_B$, and in between there is a second-order one locating at $\mu_B\simeq 390 \MeV$, which signifies the existence of CEP in the chiral phase diagram of our model.  

\begin{figure}
\centering
\includegraphics[width=75mm,clip=true,keepaspectratio=true]{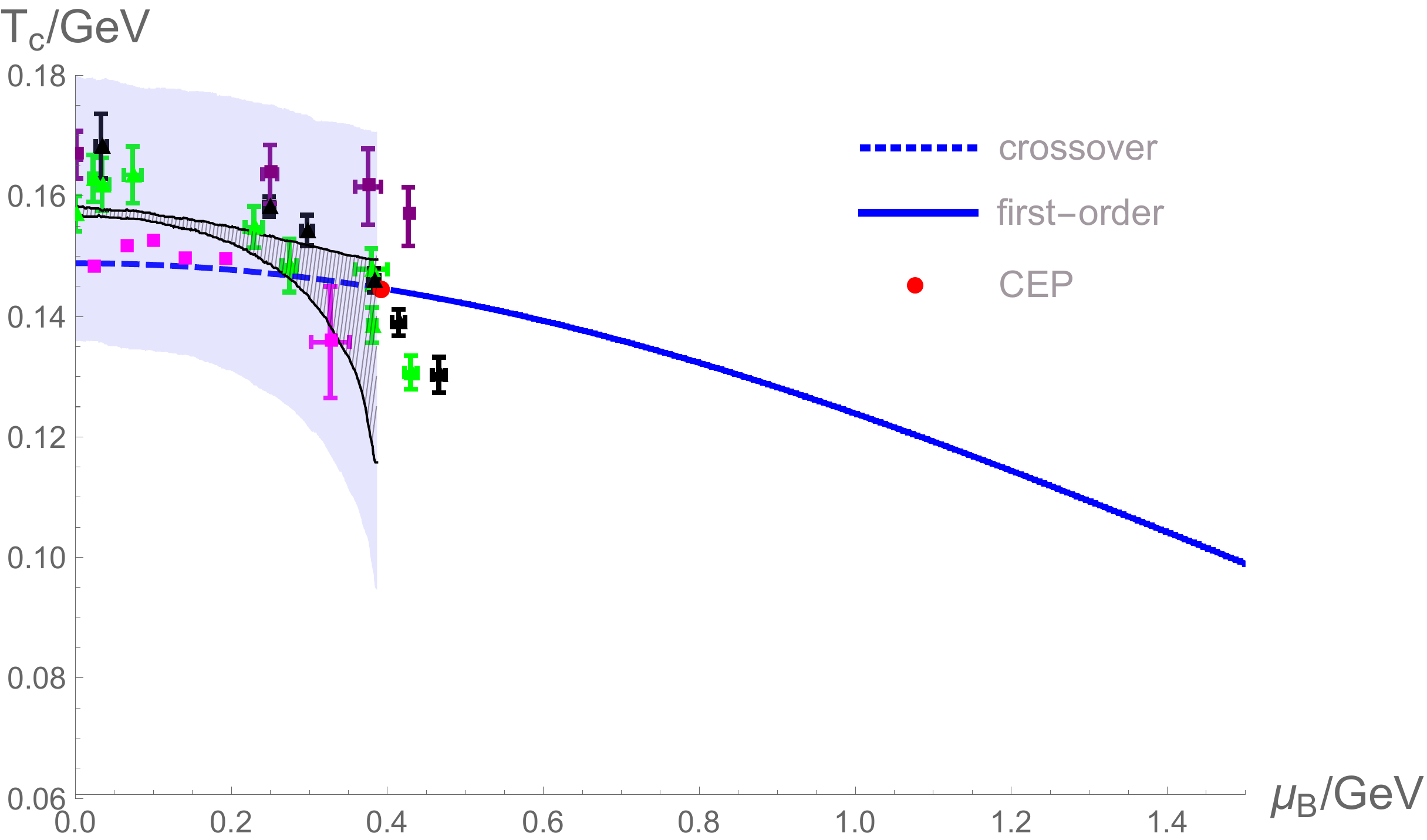}
\caption{The chiral phase diagram in the $\mu-T$ plane obtained from the improved soft-wall AdS/QCD model. The colored points with error bars are freeze-out data from experimental analysis. The black-triangle data are taken from Ref. \cite{Becattini:2005xt,Cleymans:2004pp}, the green-triangle ones are taken from Ref. \cite{Andronic:2008gu}, the purple squares are taken from Ref. \cite{Becattini:2012xb}, and the magenta squares are taken from Ref. \cite{Alba:2014eba}. The shaded black region denotes the crossover line obtained from lattice simulation at small $\mu_B$, and the light blue band indicates the uncertainty width of lattice simulation \cite{Bellwied:2015rza}.} 
\label{mu-T-diagram}
\end{figure}

We compute the chiral transition temperature $T_c$ at each baryon chemical potential $\mu_B$ using the prescription given in Ref. \cite{Fang:2018axm}. The pseudocritical temperature for the crossover transition can be defined as the maximum of $|\frac{\partial\sigma_q}{\partial T}|$, while the critical temperature for the first-order phase transition will be restricted to the region between two transition inflections of the curve. The model calculations for chiral phase diagram in the $\mu-T$ plane are shown in Fig. \ref{mu-T-diagram}, where the CEP linking the crossover transition with the first-order phase transition is highlighted by the red point with the location at $(\mu_B, T_c) \simeq (390 \MeV, 145 \MeV)$. We also show the freeze-out data analyzed from experiments of relativistic heavy ion collisions and the crossover line obtained from analytic continuation of lattice data with imaginary chemical potential \cite{Becattini:2005xt,Cleymans:2004pp,Andronic:2008gu,Becattini:2012xb,Alba:2014eba,Bellwied:2015rza}. We find that the crossover line and the location of CEP obtained from the improved soft-wall model are consistent with experimental analysis and lattice results. However, the descent of $T_c$ with the increase of $\mu_B$ at large chemical potentials is too slow for the model prediction. Thus we need to keep cautious on the reliability of the model results at large $\mu_B$.

\section{Conclusion and remarks}\label{conclution}

We have presented a further study on the improved soft-wall AdS/QCD model with $2+1$ flavors, which can reproduce the standard scenario of the quark-mass phase diagram in terms of chiral transition \cite{Fang:2018vkp}. The EOMs of octet pseudoscalar, vector and axial-vector mesons have been derived from the action of the improved soft-wall model, and the octet meson spectra were computed and also compared with experimental data. The model parameters are fixed by a global fitting with the experimental data of octet spectra. We find that the improved soft-wall model gives a better description for the pseudoscalar octet spectrum than that for the vector and axial-vector octet spectra, in which the model results for the isosinglet states are much smaller than experimental values due to the same form of EOMs for the isosinglet and isovector states in the vector octet and the tiny flavor symmetry breaking in the axial-vector octet. A possible way to address this issue is to consider high-order terms of the bulk action \cite{Sui:2010ay}. We have also computed the decay constants $f_{\pi}$, $f_{K}$, $F_{\rho}$ and $F_{a_1}$, which are compared with experimental results. 

We then studied the chemical potential effects on chiral transition, and obtained the chiral phase diagram in the $\mu -T$ plane, where the CEP still exists and locates at $(\mu_B, T_c) \simeq (390 \MeV, 145 \MeV)$. The crossover line and the location of CEP are consistent with lattice results and experimental analysis from relativistic heavy ion collisions. However, the transition temperature $T_c$ at large chemical potential declines too slowly with the increase of $\mu_B$. In our work, the baryon chemical potential $\mu_B$ and the temperature $T$ are generated by a fixed AdS/RN black hole, yet a more sensible way to introduce these effects is to consider a dynamical background which is solved from an Einstein-Maxwell-dilaton system, which also enables us to study the chiral and deconfining phase transitions simultaneously in a single holographic framework. As a preliminary attempt, we have shown that some main features of chiral phase diagram can be captured by a simply improved soft-wall AdS/QCD model, which might be taken as a guidance in further researches.

As an important aspect of QCD phase transition, we need to investigate the behaviors of equation of state that can be dealt with by the Einstein-Maxwell-dilaton system at finite chemical potential \cite{DeWolfe:2010he,Critelli:2017oub}. The back-reaction effects of the flavor sector to the background need also to be considered in order to clarify the phase structure of AdS/QCD. However, we remark that the QCD phase transition cannot be completely described by the semiclassical gauge/gravity duality even all these effects have been considered. From the large N analysis we know that the pressure of hadron gas in the confined phase is of order one, while the pressure in the deconfined phase dominated by color degrees of freedom is of order $N^2$. This must entail the string loop correction for an adequate account of thermal transition between the two phases separated by a factor of $1/N^2$ suppression in the large N limit. Moreover, there is no reason for the QCD matter produced in the heavy ion collisions to be close to thermodynamic equilibrium, therefore, it is natural for us to expect nonequilibrium effects to come into play in the phase transition with possible experimental signatures of CEP. Indeed, there have been indications that equilibrium lattice QCD susceptibilities cannot account for the moments of net-proton multiplicity distributions at low energy \cite{Borsanyi:2014ewa}, which may be a hint that the low-energy heavy ion collision at higher chemical potential is not really a thermodynamic equilibrium process. Nevertheless, we are content with a phenomenological description for low-energy QCD in terms of the bottom-up approach, which has been shown to be able to roughly grasp the picture of chiral phase transition.

\section*{Acknowledgements}
This work is partly supported by the National Natural Science Foundation of China (NSFC) under Grant Nos. 11851302, 11851303, 11747601 and 11690022, and the Strategic Priority Research Program of the Chinese Academy of Sciences under Grant No. XDB23030100 as well as the CAS Center for Excellence in Particle Physics (CCEPP). Z. F. is also supported by the NSFC under Grant No. 11905055 and the Fundamental Research Funds for the Central Universities under Grant No. 531118010198.


\bibliography{refs-AdSQCD}
\end{document}